# Quasi-classical path integral approach to supersymmetric quantum mechanics

Akira Inomata
*Department of Physics, State University of New York at Albany, Albany, New York 12222*

Georg Junker
*Institut für Theoretische Physik I, Universität Erlangen-Nürnberg, Staudtstr. 7, D-91058 Erlangen, Germany*

From Feynman's path integral, we derive quasi-classical quantization rules in supersymmetric quantum mechanics (SUSY-QM). First, we derive a SUSY counterpart of Gutzwiller's formula, from which we obtain the quantization rule of Comtet, Bandrauk and Campbell when SUSY is good. When SUSY is broken, we arrive at a new quantization formula, which is found as good as and even sometime better than the WKB formula in evaluating energy spectra for certain one-dimensional bound state problems. The wave functions in the stationary phase approximation are also derived for SUSY and broken SUSY cases. Insofar as a broken SUSY case is concerned, there are strong indications that the new quasi-classical approximation formula always overestimates the energy eigenvalues while WKB always underestimates.

## I. INTRODUCTION

The idea of supersymmetry (SUSY) is based on the expectation that there may be environments where the distinction between bosons and fermions is irrelevant [1]. Although there are some indications that the SUSY scheme works in understanding low energy phenomena [2], it is an observed fact that SUSY is generally broken. In 1981, Witten [3] utilized SUSY quantum mechanics to simulate spontaneous breaking of SUSY through non-perturbative quantum effects [4]. Despite its field theoretic origin, SUSY quantum mechanics has many interesting properties that have been utilized for solving various problems in non-relativistic quantum mechanics [5]. In fact, as early as 1976, Nicolai [6] had already used the idea of SUSY quantum mechanics for the study of statistical mechanics. As a quantization method, it has been successful in reproducing exact solutions of the Schrödinger equation for a class of so-called shape invariant potentials [7]. This relatively new method has also proven useful in studying more complex problems such as tunneling problems [8], the Kaluza-Klein monopole [9], the three body partition function for an anyon gas [10], the Pauli and the Dirac Hamiltonian for electrons in magnetic fields [11,12] and the magnetic top [13].

In 1985, Comtet, Bandrauk and Campbell [14] proposed for SUSY quantum mechanics a semiclassical quantization formula, which henceforth we shall call the CBC formula,

$$\int_{x_L}^{x_R} \sqrt{2m(\tilde{E}-\phi^2(x))}\,dx = n\pi\hbar, \qquad n = 0,1,2,\ldots, \tag{1}$$

where $x_L$ and $x_R$ are the left and right turning points given by $\phi^2(x_L) = \phi^2(x_R) = \tilde{E}$. This formula is similar to but different from the well-known WKB quantization rule,

$$\int_{x_L}^{x_R} \sqrt{2m\,(E-V(x))}\,dx = \left(n+\frac{1}{2}\right)\pi\hbar, \tag{2}$$

with $V(x_L) = V(x_R) = E$. In (1), $\tilde{E}$ is the energy shifted so that $\tilde{E} = 0$ for the ground state. It differs from $E$ of (2) by the energy shift $\epsilon$, that is, $E = \tilde{E} + \epsilon$. The real-valued function $\phi(x)$ appearing in (1), which will be referred to as the superpotential by following the recent convention, is related to the potential $V(x)$ of (2) by

$$V(x) = \phi^2(x) \pm \frac{\hbar}{\sqrt{2m}}\phi'(x) + \epsilon, \tag{3}$$

where $\phi'(x) = d\phi(x)/dx$. Depending on the sign of the second term, $E - V(x)$ equals $\tilde{E} - V_+(x)$ or $\tilde{E} - V_-(x)$ where $V_\pm(x) = \phi^2(x) \pm (\hbar/\sqrt{2m})\,\phi'(x)$. The ground state is customarily assumed to belong to $V_-(x)$, so in (1) $n = 0,1,2,\ldots$ for $V_-(x)$ and $n = 1,2,3\ldots$ for $V_+(x)$.

As is well-known, the WKB semiclassical quantization formula gives exact spectra for exactly soluble examples provided some Langer-like parameter modifications are made (see, e.g., [5]). Surprisingly, the CBC formula (1) has been found to be able to reproduce the same exact spectra without any *ad hoc* modification [14]. Even for those problems which are not exactly soluble, the CBC formula has been claimed to yield better approximation than the WKB formula does [15].

The success of the CBC formula is truly remarkable. Nevertheless, the origin of the formula is not very clear. There have been attempts to derive the CBC formula from the WKB formula. Since the WKB formula itself has the Langer-like ambiguity, such attempts only verify that whenever the WKB formula is acceptable the CBC formula should work in the same semiclassical limit. It is certainly desirable to derive the CBC formula directly from what may be taken as the first principle.

If SUSY is broken, however, the CBC formula has no ground for its validity. A natural question arises as to whether there will be a CBC counterpart for a broken SUSY case. In a previous work [16], we have proposed the following formula for a broken SUSY case,

$$\int_{x_L}^{x_R} \sqrt{2m(\tilde{E}-\phi^2(x))}\,dx = \left(n+\frac{1}{2}\right)\pi\hbar \tag{4}$$

with $\phi^2(x_L)=\phi^2(x_R)=\tilde{E}$. It has also been shown that this "broken SUSY formula" can reproduce exact spectra for the radial harmonic oscillator and the Pöschl-Teller system [17,18].

In the present paper, we wish to show that the CBC formula (1) can be derived, when SUSY is good, from Feynman's path integral by a stationary phase approximation. Then we report that the CBC counterpart, applicable only to the case where SUSY is broken, can also be derived from the same path integral. Furthermore, we propose that the new quantization formula for broken SUSY is of practical value as an approximation formula which is as good as or sometimes better than the WKB formula. However, we have no answer why the CBC formula can provide exact results for shape-invariant potentials and why the SUSY approximations whenever applicable are better than the WKB estimations. Section II briefly describes the backgrounds for our discussions. In Section III, we develop the quasi-classical approach to derive the quasi-classical counterpart of Gutzwiller's semiclassical formula for the energy-dependent Green function from which we obtain the CBC formula (1) and the broken SUSY formula (4). In Section IV, for a couple of examples with broken SUSY, including a class of power potentials, we analyze numerically the results of the WKB formula and the broken SUSY formula in comparison with those from Schrödinger's equation. For the systems whose SUSY is broken, there are indications that the standard WKB formula underestimates the energy values, while the broken SUSY formula overestimates them. A way to improve approximation for the energy spectrum of a broken SUSY system is also suggested. Remarks are made in Section V concerning the limitation of the broken SUSY formula.

## II. SUSY QUANTUM MECHANICS AND PATH INTEGRALS

Let us start with the SUSY invariant Lagrangian in one dimension [19]:

$$\begin{aligned}\mathcal{L} = &\ \frac{1}{2}m\dot{x}^2 - \phi^2(x) + \frac{i}{2}(\psi^\dagger\dot{\psi} - \dot{\psi}^\dagger\psi)\\ &-\frac{\hbar}{\sqrt{2m}}\,\phi'(x)[\psi^\dagger,\psi],\end{aligned} \tag{5}$$

where $\psi$ and $\psi^\dagger$ are Grassmann variables. From this, Feynman's path integral can formally be constructed, which consists of integrations over the bosonic coordinate variables $x$ as well as the fermionic Grassmann variables $\psi$ and $\psi^\dagger$. After carrying out path integration over the Grassmann variables [19], we can arrive at two path integrals in bosonic coordinate variables,

$$K_\pm(x'',x';\tau) = \int_{x'=x(0)}^{x''=x(\tau)} \exp\left\{\frac{i}{\hbar}\int_0^\tau L_\pm(\dot{x},x)dt\right\}\mathcal{D}x(t), \tag{6}$$

for the following effective Lagrangians,

$$L_\pm(\dot{x},x) = \frac{1}{2}m\dot{x}^2 - \phi^2(x) \mp \frac{\hbar}{\sqrt{2m}}\phi'(x). \tag{7}$$

The last term in (7) is a non-perturbative quantum effect arising from elimination of the fermionic degrees of freedom (the contribution of *all* fermion loops). The Lagrangians (7) correspond to the Hamiltonians,

$$H_\pm = -\frac{\hbar^2}{2m}\frac{d^2}{dx^2} + \phi^2(x) \pm \frac{\hbar}{\sqrt{2m}}\,\phi'(x), \tag{8}$$

which characterize the simplest version ($N=2$) of SUSY quantum mechanics as defined by Witten [3]. Naturally, the path integrals (6) describe the propagators which are associated with the time-evolution generated by the Hamiltonians (8):

$$K_\pm(x'',x';\tau) = \langle x'' \mid e^{-(i/\hbar)\tau H_\pm} \mid x'\rangle. \tag{9}$$

The Hamiltonians (8) may be factorized as

$$H_+ = AA^\dagger, \qquad H_- = A^\dagger A, \tag{10}$$

by means of the operators,

$$A = \frac{\hbar}{\sqrt{2m}}\frac{d}{dx} + \phi(x), \quad A^\dagger = -\frac{\hbar}{\sqrt{2m}}\frac{d}{dx} + \phi(x). \tag{11}$$

Therefore, it is evident that $H_\pm$ are positive semi-definite, i.e., spec$(H_\pm) \geq 0$. For the eigenvalue problems, $H_\pm\,\psi_n^{(\pm)}(x) = \tilde{E}_n^{(\pm)}\,\psi_n^{(\pm)}(x)$, we can show that $\tilde{E}_n^{(+)} = \tilde{E}_n^{(-)} = \tilde{E}_n > 0$ except for the ground state and that

$$\psi_n^{(+)}(x) = \frac{1}{\sqrt{\tilde{E}_n}}\,A\psi_n^{(-)}(x), \quad \psi_n^{(-)}(x) = \frac{1}{\sqrt{\tilde{E}_n}}\,A^\dagger\psi_n^{(+)}(x),$$

for $\tilde{E}_n \neq 0$. Customarily, the ground state is assumed to be an eigenstate of $H_-$. Namely, by choice,

$H_- \, \psi_0^{(-)}(x) = \tilde{E}_0 \, \psi_0^{(-)}(x)$. The lowest eigenvalue of $H_+$ is $\tilde{E}_1$.

If SUSY is a good symmetry, that is, if the ground state is invariant under the SUSY transformation, then

$$A \, \psi_0^{(-)}(x) = 0 \tag{12}$$

which means that there is no degeneracy in the ground state and that the ground state energy must vanish. Thus, when SUSY is a good symmetry, $\tilde{E}_0 = 0$. This relation together with (11) also implies

$$\psi_0^{(-)}(x) = \psi_0^{(-)}(x_0) \exp\left\{ -\frac{\sqrt{2m}}{\hbar} \int_{x_0}^{x} \phi(q)\, dq \right\}. \tag{13}$$

When SUSY is good, this solution must exist. In other words, the ground state function must be normalizable (or square-integrable). The normalizability of (13) poses a condition on the superpotential $\phi(x)$. The ground state function (13) is normalizable only when the integral of the superpotential $\phi(x)$ tends to infinity as $x$ goes to $\pm\infty$. For instance, when we expect a purely discrete energy spectrum, $\phi(x)$ must be such that $|\phi(x)| \to \infty$ as $|x| \to \infty$. If $\phi(x)$ is a continuous function having an odd number of zeros and the ground state (13) has a vanishing eigenvalue, $\tilde{E}_0 = 0$, then SUSY is a good symmetry [3]. On the other hand, if $\phi(x)$ is continuous and has an even number of zeros, the corresponding ground state has a strictly positive eigenvalue and will be degenerate. Hence SUSY is spontaneously broken. Therefore, the ground state of $H_-$ determines whether SUSY is broken or not. For the present purposes, it is sufficient to consider $L_-$ of (7) which corresponds to $H_-$ of (8).

Thus the path integral we wish to calculate is

$$K(x'', x'; \tau) = \int_{x'=x(0)}^{x''=x(\tau)} \exp\{(i/\hbar) S[x(t)]\} \, \mathcal{D}x(t) \tag{14}$$

with the action for $L_-$,

$$S[x(t)] = \int_0^{\tau} \left( \frac{m}{2} \, \dot{x}^2 - V_-(x) \right) dt, \tag{15}$$

where

$$V_-(x) = \phi^2(x) - \frac{\hbar}{\sqrt{2m}} \, \phi'(x). \tag{16}$$

In fact, (14) is Feynman's path integral with a potential of the form (16). However, the path integral (14), whether it is subjected to the condition (16) or not, is in general difficult to calculate. There are only a very limited number of examples that can be exactly solved by Feynman's path integral. Historically, it is known that path integration for $x \in \mathbf{R}$ can explicitly be carried out only for a quadratic system [20]. If we use polar coordinates, then we may include the harmonic oscillator in an inverse-square potential to the list of exactly soluble examples [21]. Furthermore, by mapping into a higher dimensional path integral, we can solve the Pöschl-Teller oscillator and their variations [22,23]. It is interesting to note that many of these soluble potentials are reducible to the form (3).

## III. THE QUASI-CLASSICAL APPROACH

For a more general potential system, we have to pursue an approximate solution. The most appropriate approximation method in evaluating a path integral is the stationary phase calculation. In the usual semiclassical stationary phase approximation the action functional $S[x(t)]$ is expanded about the classical path $x_{\rm cl}(t)$ as

$$S[x(t)] \simeq S_{\rm cl}(x'', x', \tau) + \delta S[x(t)] + \delta^2 S[x(t)]. \tag{17}$$

The classical action $S_{\rm cl} \equiv S_{\rm cl}(x'', x', \tau) = S[x_{\rm cl}(t)]$ is an action evaluated along the classical path from $x'$ to $x''$, determined by $\delta S[x(t)] = 0$. Then the propagator is given by the formula of Van Vleck, Pauli and Morette (VMP) [24],

$$K(x'', x'; \tau) \simeq \sum_{x_{\rm cl}(t)}^{\tau \text{ fixed}} \sqrt{\frac{i}{2\pi\hbar} \frac{\partial^2 S_{\rm cl}}{\partial x' \, \partial x''}} \, \exp\left\{ \frac{i}{\hbar} S_{\rm cl} \right\}. \tag{18}$$

If there are classical paths more than one, we have to add up the contributions from all these paths. Thus, the summation in (18) is to cover all the classical paths $x_{\rm cl}(t)$ between $x'$ and $x''$ with a fixed time interval $\tau$. This approximation formula is known to give rise to the exact propagator for the free particle, the harmonic oscillator and more general quadratic systems, but does not directly provide the WKB quantization rule (2). The energy-dependent Green function evaluated by summing over the classical paths for a fixed energy $\tilde{E}$ is the source of the WKB formula. The Fourier transformation of the semiclassical propagator (18) leads to Gutzwiller's formula [25],

$$G(x'', x'; \tilde{E}) \simeq \frac{1}{i\hbar} \frac{m}{\sqrt{|p_{\rm cl}(x') p_{\rm cl}(x'')|}} \sum_{x_{\rm cl}(t)}^{\tilde{E} \text{ fixed}} \exp\left\{ (i/\hbar) W_{\rm cl} - i(\pi/2)\nu \right\}, \tag{19}$$

where $W_{\rm cl} \equiv W_{\rm cl}(x'', x', \tilde{E}) = S_{\rm cl} + \tilde{E}\tau$ is Hamilton's characteristic function evaluated along the classical paths $x_{\rm cl}(t)$ with fixed energy $\tilde{E}$, and $\nu$ is an integral number known as the Morse index. The WKB quantization rule (2) results from the poles of this Green function. Again, the WKB formula derived by this semiclassical approximation gives the exact energy spectrum for the harmonic oscillator in one dimension. However, it does not yield the exact result for the three-dimensional radial harmonic oscillator unless the Langer replacement $\ell(\ell+1) \;\; \rightarrow \;\; (\ell+1/2)^2$ is made. For other exactly soluble examples, exact spectra may be obtained from the WKB formula by applying appropriate Langer-like modifications. In other words, most WKB results for exactly soluble systems are not exact without Langer-like replacements. The Langer replacement and other similar modifications are *ad hoc* procedures which reflect the ambiguity involved in the choice of terms of $\mathcal{O}(\hbar^2)$ for a stationary action. Applying a Langer-like replacement amounts to introducing a so-called "quantum correction" term $V_{\rm c}(x)$ in the Lagrangian. This means that an appropriate effective action,

$$S_{\rm eff}[x(t)] = \int_0^\tau \left( \frac{m}{2}\,\dot{x}^2 - V_-(x) - V_{\rm c}(x) \right) dt,$$

may be chosen so as to generate a result corrected by a Langer-like modification. So far, no general rules have been established for the choice of such correction terms. Therefore, whenever a Langer-like modification is needed, there is no inherent reason why we should make the action of the form (15) stationary. In fact, we are free in principle to make any action stationary. Since our desire is to improve the semiclassical approximation within the framework of the stationary phase approach, we must look for a more suitable effective action which may be made stationary.

### A. Sum over Quasi-Classical Paths

In order to seek an alternative and equally qualified approach, let us recall that the derivative term in (16) is the contribution from all fermion loops, while the remaining part of the action (15) contains only the contributions of the tree diagrams in perturbation calculations. It is uncertain how much such background information will remain relevant when we accept the SUSY quantum mechanical method as a tool in non-relativistic quantum mechanics. Nonetheless, we consider that the key to the alternative approach lies in separating the action (15) into two parts,

$$S[x(t)] = S_{\rm tree}[x(t)] + \hbar\varphi[x(t)], \qquad (20)$$

with the "tree" action,

$$S_{\rm tree}[x(t)] = \int_0^\tau \left( \frac{m}{2}\,\dot{x}^2 - \phi^2(x) \right) dt, \qquad (21)$$

and the "fermionic" action representing the fermion loop corrections,

$$\varphi[x(t)] = \frac{1}{\sqrt{2m}} \int_0^\tau \phi'(x(t))\, dt. \qquad (22)$$

Then, we pursue an approximation for which the tree action (21) is made stationary.

Namely, we demand that

$$\delta S_{\rm tree}[q(t)] = 0 \qquad (23)$$

which determines specific paths $x = q(t)$ with momentum

$$p_{\rm qc}(q) = m\dot{q} = \pm\sqrt{2m(E - \phi^2(q))}, \qquad (24)$$

and energy $E = -\partial S_{\rm tree}[q(t)]/\partial\tau$. Here, the $+$ and $-$ signs are to be taken for the motion to the right and left, respectively. Since the "classical" paths $x_{\rm cl}(t)$ are specified by $\delta S[x(t)] = 0$, the paths $q(t)$ along which the tree action (21) is stationary are not quite classical. We shall call them "quasi-classical" paths, and we shall carry out the quasi-classical stationary phase calculation[1] for the path integral (14).

Now we expand the tree action (21) about a quasi-classical path $q(t)$ to second order by letting $\xi(t) = x(t) - q(t)$ with $x(0) = q(0) = x_{\rm cl}(0) = x'$ and $x(\tau) = q(\tau) = x_{\rm cl}(\tau) = x''$:

$$S_{\rm tree}[x(t)] \simeq S_{\rm qc}(x'', x', \tau) + \delta^2\tilde{S}[\xi(t)]\,. \qquad (25)$$

Here, $S_{\rm qc} \equiv S_{\rm qc}(x'', x', \tau) = S_{\rm tree}[q(t)]$ is the tree action (21) evaluated along the quasi-classical path $q(t)$, and the second variation $\delta^2\tilde{S}[\xi(t)]$, representing the contribution from the fluctuation about the quasi-classical paths, is given by

$$\delta^2\tilde{S}[\xi(t)] = \int_0^\tau \left( \frac{1}{2}\,m\dot{\xi}^2 - \frac{1}{2}\,m\Omega^2(t)\,\xi^2 \right) dt \qquad (26)$$

where $m\,\Omega^2(t) = d^2\phi^2(q(t))/dq^2$.

[1] In the literature the terms "semiclassical" and "quasi-classical" are often synonymous. Here, in contrast, we are differentiating the quasi-classical calculation from the usual semiclassical approximation.

Then we calculate the path integral (14) by the stationary phase approximation. Because of the quasi-classical path dominance, the fermionic action (22) takes the form,

$$\varphi \equiv \varphi[q(t)] = \frac{1}{2} \int_{\phi(x')}^{\phi(x'')} \frac{d\phi}{\pm\sqrt{E-\phi^2}} \tag{27}$$

which immediately integrates, if no turning points are involved, to

$$\varphi = \tfrac{1}{2}\left(a(x'') - a(x')\right) \tag{28}$$

where

$$a(q) = \arcsin \frac{\phi(q)}{\sqrt{E}} \,. \tag{29}$$

In general, the quasi-classical motion takes place periodically between the left turning point $x_L$ and the right turning point $x_R$ of the quasi-classical paths. At these turning points, $p_{\rm qc}(x_L) = p_{\rm qc}(x_R) = 0$, and the momentum changes its sign. The fermionic action (22) thus turns to an overall phase to the partial propagator corresponding to each quasi-classical path:

$$\varphi_k = \varphi_0 + k[a(x_R) - a(x_L)] \,, \tag{30}$$

where $k$ signifies the number of full cycles of the path $q(t)$, and $\varphi_0$ is the remaining phase due to incomplete cycles. If, in particular, the path has no turning point, the phase takes the special form (28).

The quasi-classical approximation of the action (20)

$$S[x(t)] \simeq S_{\rm qc} + \hbar\varphi + \delta^2 \tilde{S}[\xi(t)]$$

reduces the path integral (14) to

$$K(x'', x'; \tau) \simeq \sum_{q(t)}^{\tau \text{ fixed}} K(0,0;\tau) \exp\left\{\frac{i}{\hbar} S_{\rm qc} + i\varphi\right\} \,, \tag{31}$$

where

$$K(0,0;\tau) = \int_{\xi(0)=0}^{\xi(\tau)=0} \exp\left\{\frac{i}{\hbar}\delta^2 \tilde{S}[\xi(t)]\right\} \mathcal{D}\xi(t) \,. \tag{32}$$

The remaining path integral (32) for the second variation (26) over the $\xi$-variable can easily be calculated by the standard technique [26], which results in the prefactor of the form,

$$K(0,0;\tau) = \sqrt{\frac{m}{2\pi i\hbar f(\tau)}} \,. \tag{33}$$

The function $f(t)$ appearing in this prefactor is the solution of

$$\left[\frac{d^2}{dt^2} + \Omega^2(t)\right] f(t) = 0, \tag{34}$$

which satisfies the initial conditions, $f(0) = 0$ and $\dot{f}(0) = 1$. The solution can be put, as is easily checked, in the form (see, e.g., [27]),

$$f(\tau) = \dot{q}(\tau)\dot{q}(0) \int_0^{\tau} \frac{dt}{[\dot{q}(t)]^2} \,. \tag{35}$$

Since the velocity $\dot{q}(t)$ changes sign at every turning point, we have the relation,

$$m^2 \dot{q}(\tau)\dot{q}(0) = (-1)^{\nu} |p_{\rm qc}(x'')p_{\rm qc}(x')| \,. \tag{36}$$

Here $\nu$ is the number of turning points along a particular quasi-classical path $q(t)$ between $x'$ and $x''$, that is, the number of the zeros of $f(t)$ for $t \in [0,\tau]$, which is in this case the same as the Morse index [28]. It is zero when $\tau$ is small, and hence the solution $f(\tau)$ of (34) satisfies the desired initial conditions.

The right hand side of (35) may also be expressed in different fashions. Integrating the inverse of (24) gives the time interval $\tau$ as a function of $x'$, $x''$ and $E$,

$$\tau = \int_{x'}^{x''} \frac{dq}{\dot{q}} = m \int_{x'}^{x''} \frac{dq}{\pm\sqrt{2m(E-\phi^2(q))}} \,. \tag{37}$$

In the last equality, the negative sign is to be chosen when $\dot{q} < 0$. Differentiating both sides of this yields

$$\left(\frac{\partial \tau}{\partial E}\right) = -\frac{1}{m} \int_0^{\tau} \frac{dt}{[\dot{q}(t)]^2} \,. \tag{38}$$

Combining (36) and (38), we may rewrite (35) as

$$f(\tau) = -\frac{1}{m} (-1)^{\nu} \left|p_{\rm qc}(x'')p_{\rm qc}(x')\right| \frac{\partial \tau}{\partial E} \,. \tag{39}$$

If we further define the quasi-classical counterpart of Hamilton's characteristic function,

$$W_{\rm qc}(x'', x', E) = S_{\rm qc}(x'', x', \tau) + E\tau = \int_{x'}^{x''} p_{\rm qc}(q)\, dq, \tag{40}$$

we get

$$\frac{\partial W_{\rm qc}}{\partial E} = \tau, \qquad \frac{\partial^2 W_{\rm qc}}{\partial E^2} = \frac{\partial \tau}{\partial E} \,, \tag{41}$$

as well as

$$\begin{aligned} p_{\rm qc}(x'') &= \frac{\partial W_{\rm qc}}{\partial x''} = \frac{\partial S_{\rm qc}}{\partial x''}, \\ p_{\rm qc}(x') &= -\frac{\partial W_{\rm qc}}{\partial x'} = -\frac{\partial S_{\rm qc}}{\partial x'} \,. \end{aligned} \tag{42}$$

At this point, if we wish, we can derive the quasi-classical counterpart of the VMP formula (18) in a way parallel to the standard derivation of (18). From (42) it is obvious that

$$\frac{\partial^2 S_{\rm qc}}{\partial x' \partial x''} = \frac{\partial p_{\rm qc}(x'')}{\partial x'} = \frac{m}{p_{\rm qc}(x'')} \frac{\partial E}{\partial x'} ,$$

where we have used the relation $\partial p_{\rm qc}(q)/\partial E = m/p_{\rm qc}(q)$ resulting from (24). From the Hamilton-Jacobi equation for the quasi-classical paths, $E = -\partial S_{\rm qc}/\partial \tau$, there follows

$$\frac{\partial E}{\partial x'} = -\frac{\partial^2 S_{\rm qc}}{\partial \tau \partial x'} = \frac{\partial p_{\rm qc}(x')}{\partial \tau} = \frac{m}{p_{\rm qc}(x')} \left( \frac{\partial \tau}{\partial E} \right)^{-1} .$$

Using these results, we replace $\partial\tau/\partial E$ in (38) by

$$\left( \frac{\partial \tau}{\partial E} \right)^{-1} = \frac{1}{m^2} p_{\rm qc}(x') p_{\rm qc}(x'') \frac{\partial^2 S_{\rm qc}}{\partial x' \partial x''}$$

and arrive at a quasi-classical counterpart of the VPM formula,

$$K(x'', x'; \tau) \simeq \sum_{q(t)}^{\text{fixed } \tau} \sqrt{\frac{i}{2\pi\hbar} \left| \frac{\partial^2 S_{\rm qc}}{\partial x' \, \partial x''} \right|} \times \exp\left\{ \frac{i}{\hbar} S_{\rm qc} + i\varphi - i\frac{\pi}{2}\nu \right\} . \tag{43}$$

### B. Gutzwiller's Formula for SUSY-QM

To find the quasi-classical counterpart of Gutzwiller's formula (19), we have to go over from the propagator to the energy-dependent Green function by the Fourier transformation,

$$G(x'', x'; \tilde{E}) = \frac{1}{i\hbar} \int_0^\infty K(x'', x'; \tau)\, e^{(i/\hbar)\tilde{E}\tau}\, d\tau. \tag{44}$$

In going over to Gutzwiller's semiclassical formula (19), the VPM formula (18) is usually used [26,29]. In much the same manner, we may utilize the quasi-classical counterpart (43) of the VPM formula to compute the time-integral of (44). However, for clarity, we choose a slightly different approach.

Substituting (39) and (41) into (33), we express the propagator (31) in the form,

$$K(x'', x'; \tau) \simeq \sum_{q(t)}^{\tau \text{ fixed}} \sqrt{\frac{1}{2\pi i\hbar\, (-1)^\nu |p_{\rm qc}(x') p_{\rm qc}(x'')|}} \left| \frac{\partial^2 W_{\rm qc}}{\partial E^2} \right|^{-1/2} \exp\left\{ \frac{i}{\hbar} [W_{\rm qc}(E) - E\tau] + i\varphi \right\} . \tag{45}$$

Then we put this into (44) and convert the $\tau$-integration into the $E$-integration by letting $d\tau = |\partial\tau/\partial E|\, dE = |\partial^2 W_{\rm qc}/\partial E^2|\, dE$ to get

$$G(x'', x'; \tilde{E}) \simeq \frac{i}{\hbar} \int \sum_{q(t)} \sqrt{\frac{m}{2\pi i\hbar (-1)^\nu |p_{\rm qc}(x') p_{\rm qc}(x'')|}} \left| \frac{\partial^2 W_{\rm qc}}{\partial E^2} \right|^{1/2} \exp\left\{ \frac{i}{\hbar} W_{\rm qc}(E) + \frac{i}{\hbar} (\tilde{E} - E)\, \tau + i\varphi \right\} dE. \tag{46}$$

Next, we expand the function $W_{\rm qc}(E)$ about $\tilde{E}$ as

$$W_{\rm qc}(E) = W_{\rm qc}(\tilde{E}) + \left( \frac{\partial W_{\rm qc}}{\partial E} \right)_{\tilde{E}} (E - \tilde{E}) + \frac{1}{2} \left( \frac{\partial^2 W_{\rm qc}}{\partial E^2} \right)_{\tilde{E}} (E - \tilde{E})^2 + \cdots \tag{47}$$

and perform the stationary phase integration over $E$ about $\tilde{E}$. In this manner, we are able to arrive at the following quasi-classical approximation formula for the energy-dependent Green function,

$$G(x'', x'; \tilde{E}) \simeq \frac{1}{i\hbar} \frac{m}{\sqrt{|p_{\rm qc}(x') p_{\rm qc}(x'')|}} \sum_{q(t)}^{\tilde{E} \text{ fixed}} \exp\left\{ (i/\hbar) W_{\rm qc}(\tilde{E}) + i\varphi - i(\pi/2)\nu \right\} \tag{48}$$

where the sum is over all quasi-classical paths $q(t)$ from $x'$ to $x''$ with a fixed energy $\tilde{E}$. This is the quasi-classical counterpart of Gutzwiller's formula (19).

### C. Quasi-Classical Quantization

For each quasi-classical trajectory with the same fixed energy value $\tilde{E}$, we have to find $W_{\rm qc}$, $\varphi$ and $\nu$ in order to perform the summation in (48). Following Schulman's prescription [26], we group the set of all paths into four classes:

(1) paths which leave $x'$ to the right and arrive at $x''$ from the left;

(2) those which leave $x'$ to the right and arrive at $x''$ from the right;

(3) those which leave $x'$ to the left and arrive at $x''$ from the left; and

(4) those which leave $x'$ to the left and arrive at $x''$ from the right.

They may be reflected at the left turning point $x_L$, or the right turning point $x_R$, or at the left and right turning points. Within each class, denoted by the superscript $i\,(i=1,2,3,4)$ in parentheses, a path is uniquely characterized by a non-negative integer $k$ indicating the number of full cycles the path contains. For a path with $k$ cycles in class $(i)$ $(i=1,2,3,4)$, we write $(W_{\rm qc}^{(i)} \equiv W_k^{(i)})$

$$\begin{aligned} W_k^{(i)} &= W_0^{(i)} + 2kw(x_R), \\ \varphi_k^{(i)} &= \varphi_0^{(i)} + k\left[a(x_R) - a(x_L)\right], \\ \nu_k^{(i)} &= \nu_0^{(i)} + 2k, \end{aligned} \tag{49}$$

where

$$w(x) = \int_{x_L}^{x} \sqrt{2m(\tilde{E} - \phi^2(q))}\, dq\,, \tag{50}$$

and $a(x)=$arcsin$\left(\phi(x)/\sqrt{\tilde{E}}\right)$ as given in (29). The quantities defined for $k=0$ are

$$\begin{aligned} W_0^{(1)} &= w(x'') - w(x'), \\ \varphi_0^{(1)} &= \tfrac{1}{2}\left[a(x'') - a(x')\right], \\ \nu_0^{(1)} &= 0, \end{aligned}$$

$$\begin{aligned} W_0^{(2)} &= w(x'') + w(x'), \\ \varphi_0^{(2)} &= \tfrac{1}{2}\left[a(x'') + a(x')\right] - a(x_L), \\ \nu_0^{(2)} &= 1, \end{aligned}$$

$$\begin{aligned} W_0^{(3)} &= 2w(x_R) - w(x'') - w(x'), \\ \varphi_0^{(3)} &= a(x_R) - \tfrac{1}{2}\left[a(x'') - a(x')\right], \\ \nu_0^{(3)} &= 1, \end{aligned}$$

and

$$\begin{aligned} W_0^{(4)} &= 2w(x_R) - w(x'') + w(x'), \\ \varphi_0^{(4)} &= a(x_R) - a(x_L) - \tfrac{1}{2}\left[a(x'') - a(x')\right], \\ \nu_0^{(4)} &= 2. \end{aligned}$$

Now the sum for a fixed $\tilde{E}$ in (48) can be rewritten as

$$\sum_{q(t)}^{\tilde{E}\,\text{fixed}} (\cdot) = \sum_{i=1}^{4}\sum_{k=0}^{\infty}(\cdot).$$

The second summation leads to a geometric series which can be easily evaluated. As a result, the Green function in the quasi-classical approximation becomes

$$G(x'',x';\tilde{E}) \simeq \frac{m}{i\hbar\sqrt{|p_{\rm qc}(x')p_{\rm qc}(x'')|}}\,\frac{\sum_{i=1}^{4}\exp\left\{(i/\hbar)W_0^{(i)} + i\varphi_0^{(i)} - i(\pi/2)\nu_0^{(i)}\right\}}{1-\exp\left\{i\left[2w(x_R)/\hbar + a(x_R) - a(x_L) - \pi\right]\right\}}. \tag{51}$$

Apparently, the poles of the Green function (51) occur when

$$w(x_R) = \left(n + \frac{1}{2} - \frac{a(x_R)-a(x_L)}{2\pi}\right)\pi\hbar \tag{52}$$

where $n=0,1,2,\ldots$. This is indeed a quasi-classical quantization condition, which implies two different formulas depending on whether SUSY is good or broken. To determine the values of $a(x_L)$ and $a(x_R)$ explicitly, we have to recall the conditions at the two turning points, $\phi^2(x_L)=\phi^2(x_R)=\tilde{E}$, or $|\phi(x_L)|=|\phi(x_R)|=\sqrt{\tilde{E}}$, which has two solutions [16]. Accordingly, we have the following two distinct cases:

*Case I:* $\phi(x_R) = -\phi(x_L) = \pm\sqrt{\tilde{E}}$. In this case, it is apparent from (24) that $a(x_R) = -a(x_L) = \pm \arcsin 1 = \pm\pi/2$. Since the ground state is assumed to be an eigenstate of $H_-$, the upper sign should be selected.[2] The fermionic phase $\varphi$ has a non-zero contribution which cancels the fractional part on the right hand side of (52). Consequently, the quantization condition (52) results in

$$w(x_R) = n\pi\hbar, \qquad n = 0, 1, 2, \ldots, \tag{53}$$

which coincides with the celebrated formula (1) of Comtet, Bandrauk and Campbell [14]. Since $\phi(x)$ must have an odd number of zeros in this case, SUSY is a good symmetry. For the residues of the Green function $G(x'', x'; \tilde{E})$ at the poles due to (53), we have

$$\operatorname{Res} G(x'', x'; \tilde{E})\Big|_{\tilde{E}=\tilde{E}_n} \simeq \frac{4m}{\tau_n} \frac{\cos\left(\frac{1}{\hbar}\, w(x') + \frac{1}{2}\, a(x')\right)}{\sqrt{|p_{\rm cl}(x')|}} \frac{\cos\left(\frac{1}{\hbar}\, w(x'') + \frac{1}{2}\, a(x'')\right)}{\sqrt{|p_{\rm cl}(x'')|}}, \tag{54}$$

where

$$\tau_n = \int_{x_L}^{x_R} \sqrt{2m/(\tilde{E}_n - \phi^2(q))}\, dq \tag{55}$$

is the period (the time interval for a single cycle) of the bounded motion with energy $\tilde{E}_n$. From this follow the excited state wave functions for $q_L < x < q_R$,

$$\psi_n^{(\pm)}(x) \simeq \sqrt{\frac{4m}{\tau_n}}\, |p_{\rm cl}(x)|^{-1/2} \cos\left(\frac{w(x)}{\hbar} \mp \frac{1}{2} \arcsin \frac{\phi(x)}{\sqrt{\tilde{E}_n}}\right), \tag{56}$$

which is normalized within the quasi-classical scheme. In the path integral approach, the results are usually normalized by the basic requirement,

$$\lim_{\tau \to 0} K(x'', x'; \tau) = \delta(x'' - x').$$

Note that these wave functions are only for the excited states with $\tilde{E}_n > 0$ $(n = 1, 2, 3, \ldots)$. The ground state wave function has already been given by (13).

*Case II:* $\phi(x_R) = \phi(x_L) = \pm\sqrt{\tilde{E}}$. In this case, $a(x_R) = a(x_L) = \pm \arcsin 1 = \pm\pi/2$ and the contribution of the fermionic phase $\varphi$ vanishes in (52). Therefore, the quantization condition (52) implies another formula,

$$w(x_R) = \left(n + \frac{1}{2}\right) \pi\hbar\ , \qquad n = 0, 1, 2, \ldots, \tag{57}$$

which is identical with the formula (4) we have proposed for the broken SUSY case. The left-hand side of this formula is the same as that of the CBC formula (1), whereas the right-hand side equals that of the WKB formula (2). This new formula, which is a hybrid of the WKB and the CBC formula, is nothing but the usual WKB result for the "tree" Hamiltonian without the fermion loop contributions ($\varphi = 0$). Since $\phi(x)$ has an even number of zeros, SUSY is broken. The hybrid formula (57) is indeed the CBC counterpart for broken SUSY cases. Although the fermion loops do not affect the spectrum in the present approximation, there is an effect on the quasi-classical eigenfunctions. The residues of the Green function at the poles due to (57) are

$$\operatorname{Res} G(x'', x'; \tilde{E})\Big|_{\tilde{E}=\tilde{E}_n} \simeq \frac{4m}{\tau_n} \frac{\sin\left(\frac{1}{\hbar}\, w(x') + \frac{1}{2}\, a(x')\right)}{\sqrt{|p_{\rm cl}(x')|}} \frac{\sin\left(\frac{1}{\hbar}\, w(x'') + \frac{1}{2}\, a(x'')\right)}{\sqrt{|p_{\rm cl}(x'')|}}. \tag{58}$$

We immediately find the quasi-classical wave functions,

[2] To obtain the quasi-classical spectrum for $H_+$ the lower sign has to be chosen. This leads to the same quantization condition (53). However, $n = 1, 2, 3, \ldots$ for this case.

$$\psi_n^{(\pm)}(x) \simeq \sqrt{\frac{4m}{\tau_n}}\, |p_{\rm qc}(x)|^{-1/2} \sin\left(\frac{w(x)}{\hbar} \mp \frac{1}{2}\arcsin\frac{\phi(x)}{\sqrt{\tilde{E}_n}}\right), \tag{59}$$

where $n = 0, 1, 2, \ldots$, and $q_L < x < q_R$.

The quasi-classical quantization condition (52) derived in the above from the path integral is identical with the rule obtained earlier by Eckhardt [31] from Maslov's asymptotic analysis. In fact, he also identified the first case (53) with the CBC formula. However, he did not recognized that the second case (57) corresponds to broken SUSY. The hybrid formula (57) was for the first time related to broken SUSY in ref. [16] and applied to solve quantum mechanical problems in ref. [17,18].

## IV. APPLICATIONS OF THE BROKEN SUSY FORMULA

### A. Exactly Soluble Examples

It is a remarkable fact that the CBC quantization rule (1) supplies exact energy spectra for shape invariant potentials while the WKB rule (2) needs the Langer-like modification. What may as well be remarkable is that the broken SUSY formula (57) can also provide without any *ad hoc* modification the exact spectra of systems for which the superpotential has a general form [17,18],

$$\phi(x) = Af(x) + B/f(x), \tag{60}$$

where $A$ and $B$ are constants. The radial harmonic oscillator, the Pöschl-Teller oscillator, and the modified Pöschl-Teller oscillator are such examples [17,18]. Since these examples are basic, it would be instructive to compare SUSY compatible forms and broken SUSY compatible forms of the superpotential explicitly.

(a) *The radial harmonic oscillator*:

$$V_-(r) = \frac{1}{2}m\omega^2 r^2 + \frac{l(l+1)\hbar^2}{2mr^2} + (\eta - 1/2)\hbar\omega\,,$$

($l = 0, 1, 2, \ldots$, $r > 0$) for which

$$\phi(x) = \sqrt{\frac{m}{2}}\omega r + \eta\frac{\hbar}{\sqrt{2m}\, r},$$

where $\eta = -l - 1$ (good SUSY) or $\eta = l$ (broken SUSY).

(b) *The Pöschl-Teller oscillator*:

$$V_-(x) = V_0[\kappa(\kappa-1)\csc^2 x + \lambda(\lambda-1)\sec^2 x] - V_0(\eta - \lambda)^2\,,$$

($\kappa > 1$, $\lambda > 1$, $0 < x < \pi/2$) for which

$$\phi(x) = \sqrt{V_0}[\lambda \tan x + \eta \cot x],$$

where $\eta = -\kappa$ (good SUSY) or $\eta = \kappa - 1$ (broken SUSY).

(c) *The modified Pöschl-Teller oscillator*:

$$V_-(x) = V_0[\kappa(\kappa-1)\,\mathrm{csch}^2 x - \lambda(\lambda-1)\,\mathrm{sech}^2 x] - V_0(\eta - \lambda)^2\,,$$

($\kappa > 1$, $\lambda > 1$, $x > 0$) for which

$$\phi(x) = \frac{\hbar}{\sqrt{2m}}[\lambda\tanh(x) + \eta\coth(x)],$$

where $\eta = -\kappa$ (good SUSY) or $\eta = \kappa - 1$ (broken SUSY).

There are other systems soluble exactly by the broken SUSY formula [32], but those soluble are reducible to the form given above.

### B. The Power Potential Problems

The new quasi-classical formula (4), like the WKB formula (2) and the CBC formula (1), can serve significantly as an approximation formula even for non-shape invariant systems. In order to present a conspicuous difference between the approximate results of a SUSY system and a broken-SUSY system, let us studies a few examples.

The first example we consider is a class of superpotentials of the form,

$$\phi(x) = Ax^k\ ,\quad A > 0\ ,\quad k = 1, 2, 3, \ldots\ , \tag{61}$$

which is not shape-invariant and not exactly solvable. In a recent work [15], Khare has studied the case of the odd power, that is, the case of $k = 2d-1$ where $d$ is a positive integer. Since the superpotential is of an odd power, it has an a single zero but a zero of order $2d-1$. Therefore, the CBC formula is applicable. In the case of an even power, say, $k = 2d$, the zero of the superpotential is of order $2d$, so that SUSY is broken. The CBC formula is no longer applicable. In the earlier work [15], the even power case has been left out from being treated within the framework of SUSY quantum mechanics. The new broken SUSY formula (4) enables us to handle even power cases and to supplement Khare's analysis of the odd power problem.[3]

The partner potentials derived from (61) for $k = 2d$ are

$$V_{\pm}(x) = A^2 x^{4d} \pm \frac{2\hbar A\, d}{\sqrt{2m}}\, x^{2d-1}. \qquad (62)$$

Obviously, $V_+(x) = V_-(-x)$. Hence it is sufficient to consider $V(x) = V_-(x) + \epsilon$. As is the case of the CBC integral for an odd power [15], the integral in the new broken SUSY approximation formula (4) can be explicitly calculated, yielding an approximate energy spectrum expressed in a closed form,

$$\begin{aligned}\tilde{E}_n \;=\; & A^{2/(2d+1)} \left(\frac{\hbar^2}{2m}\right)^{2d/(2d+1)} \\ & \times \left[\frac{\Gamma\left(\frac{6d+1}{4d}\right)}{\Gamma\left(\frac{4d+1}{4d}\right)}\sqrt{\pi}\left(n+\frac{1}{2}\right)\right]^{4d/(2d+1)} .\end{aligned} \qquad (63)$$

We compare numerically these energy values with those obtained from the usual WKB approximation (2). In addition, we make exact numerical integrations of the Schrödinger equation. The relative deviations $\Delta = (E_{\mathrm{exact}} - E_{\mathrm{approx}})/E_{\mathrm{exact}}$ (in %) from the exact eigenvalues are shown graphically in Figure 1 for $d = 1$, 2 and 3. We have chosen units such that $A = m = \hbar = 1$. For the ground state energy shift we have taken $\epsilon = 2$. Graphs of the corresponding potentials $V_-(x)$ are given in Figure 2.

The results for $d = 1$ show that the WKB approximation of the ground-state energy ($n = 0$) is better than the estimate by formula (4). Even for excited states the results of WKB are generally better than those from (4). However, for the cases where $d = 2$ and 3, the broken SUSY formula (4) provides better approximations for the ground-state energies. Except for the first few exited states, formula (4) is always better than WKB. In the limit $d \to \infty$, we obtain from (63)

$$\lim_{d\to\infty} \tilde{E}_n = (\hbar^2\pi^2/32m)\,(2n+1)^2 \qquad (64)$$

which coincides with the energy spectrum for the even states of a particle in an infinite square well of width 4 ($-1 < x < 3$). In fact, for large $d$, the potential $V_-(x)$ has the form of an infinite square well of width 2 with an additional deep negative dip near $x = 1$. The area enclosed by the negative part of $V_-(x)$ and the positive $x$-axis diverges as $d \to \infty$. This is in contrast to the odd power potential considered by Khare, which reduces in the case of $V_+(x)$ to a standard one-dimensional box in the limit $d \to \infty$.

The energy eigenfunctions for large $d$, as observed by numerical integrations, are of the form

$$\psi_n(x) = N\, F(x)\, \sin\left[(2n+1)(x+1)\pi/4\right] \qquad (65)$$

where $N$ is a normalization constant and the envelope function $F(x)$ has for large $d$ a shape as shown in Figure 3. In the limit $d \to \infty$ it consists of two unit-step functions denoted by $\theta$:

$$\lim_{d\to\infty} F(x) = \frac{1}{2}\Big(\theta(x+1) - \theta(x-1)\Big). \qquad (66)$$

This means that the potential $V_-(x)$ becomes that of an infinite square well with Dirichlet boundary conditions at $x = -1$ and Neumann boundary conditions at $x = 1$. It should be noted that the behavior of $V_-(x)$ near $x = 1$ is very similar to that of the derivative-delta potential $U(x) = -\beta\delta'(x-1)$ where $\beta \to \infty$ as $d \to \infty$. It is known that $U(x)$ for $\beta \to \infty$ produces the Neumann condition at $x = 1$ and decouples the two half-lines $x < 1$ and $x > 1$ [33]. These asymmetric boundary conditions are in a sense remnants of the broken SUSY. Therefore, in the limit $d \to \infty$, the potential $V(x)$ becomes equivalent to the left half ($-1 < x < 1$) of an infinite square well for $-1 < x < 3$ having the $\delta'$-potential at $x = 1$, in which only the even states of the standard infinite square well of width 4 survive. The spectrum (62) is indeed exact in the limit $d \to \infty$. This would suggest that for large $d$ the new formula (4) will give better energy estimates than the WKB formula (2) does.

Although the broken SUSY formula (4) does not necessarily improve approximations as is seen in the case of $d = 1$, it adds an important advantage if applied in combination with the WKB formula (2). A closer look at the relative errors given in Figure 1 makes us realize that the WKB estimates are always below the exact values and that the results of the broken SUSY formula always above the exact values. The mean values of the results of (2) and (4) will give improved energy estimation. This interesting feature is not unique to the power potentials. There are strong indications that for any broken SUSY system the WKB result is an underestimation whereas the broken SUSY calculation results in an overestimation.

### C. More Examples and Improved Approximations

Let us examine additional examples for which the superpotentials are not power functions:

(a) $\phi(x) = \cosh x$ and

[3] After the first version of this analysis was submitted for publication, a similar work on the even power potential appeared in ref. [30]. The emphasis there differs from ours.

(b) $\phi(x) = \exp(x^2/2)$.

Both of these superpotentials satisfy the symmetric conditions $\phi(x_L) = \phi(x_R) = \sqrt{\tilde{E}}$ at the turning points. Naturally, the CBC formula is inapplicable. Thus, we calculate the energy values by using the WKB formula and the broken SUSY formula, and compare their results. Here, for convenience, the ground state energy shift has been set to $\epsilon = 0$, which implies $E = \tilde{E}$. In Figure 4, we observe again that the WKB formula (2) underestimates the energy values whereas the broken SUSY formula (4) overestimates them. We have also examined other examples [18,34]. Remarkably, *all* the examples we have examined so far have commonly exhibited the features that the results of the broken SUSY formula are as good as the WKB approximations and that the two estimations are separately distributed above and below the exact values. Therefore, averaging the results from the two formulas will lead to an improved approximation for a system whose superpotential satisfies the condition $\phi(x_L) = \phi(x_R) = \pm\sqrt{E}$:

$$E_{\rm imp} = \tfrac{1}{2}\left(E_{\rm WKB} + E_{\rm broken}\right). \qquad (67)$$

In Table I and Table II, numerical comparisons between the improved values and the exact values are shown for the examples (a) and (b). In both cases the broken SUSY formula is better than the WKB formula. This is clearly demonstrated by the negative sign before the deviations of the improved values from the exact results. Hence averaging of the WKB and the CBC results has no particular significance. In Figure 5, we show a numerical comparison of the WKB estimation and the CBC approximation for the odd potentials $\phi(x) = x|x|$, $\phi(x) = x^3$ and $\phi(x) = x^5$. This time, we employ the shift $\epsilon = 2$. Naturally, the WKB formula (2) yields different results for the two potentials $V_-(x)$ and $V_+(x)$. In Figure 5, although we use the same symbols for both WKB results, the data for $V_+(x)$ are connected by thick lines while those for $V_-(x)$ are connected by thin lines. For the CBC formula (1) or (52), we have to use by choice only $V_-(x)$ to which the ground state belongs. Note that the lowest energy for $V_+(x)$ is $\tilde{E}_1$. Since the ground state energy of $V_-(x)$ vanishes in this case, the relative deviations $\Delta$ become infinite. Therefore, the estimates for the ground state energy by the WKB formula are omitted in Figure 5. As is apparent in Figure 5, the WKB calculation for $V_-$ sometimes overestimates and the CBC value may become lower than the Schrödinger evaluation. The CBC approximation for $\phi(x) = x|x|$ with $n = 2$ gives $\Delta = -0.009\%$ and thus is slightly below the exact value.

TABLE I. Comparison for $\phi(x) = \cosh\, x$. The relative errors parenthesized are given in % .

| $n$ | exact | WKB | broken SUSY | improved values[a] |
|---|---|---|---|---|
| 0 | 1.7190 | 1.6688 | 1.7650 | 1.7169 |
| | | (−2.92) | (+2.68) | (−0.12) |
| 1 | 3.5578 | 3.5165 | 3.5890 | 3.5528 |
| | | (−1.16) | (+0.88) | (−0.14) |
| 2 | 5.7122 | 5.6754 | 5.7380 | 5.7067 |
| | | (−0.64) | (+0.45) | (−0.10) |
| 3 | 8.1421 | 8.1088 | 8.1652 | 8.1373 |
| | | (−0.41) | (+0.28) | (−0.06) |
| 4 | 10.8210 | 10.7906 | 10.8430 | 10.8168 |
| | | (−0.28) | (+0.20) | (−0.04) |
| 5 | 13.7291 | 13.7021 | 13.7516 | 13.7269 |
| | | (−0.20) | (+0.16) | (−0.02) |

[a] Obtained by averaging the WKB and broken SUSY result according to (67).

TABLE II. Comparison for $\phi(x) = \exp(x^2/2)$. The relative errors parenthesized are given in % .

| $n$ | exact | WKB | broken SUSY | improved values |
|---|---|---|---|---|
| 0 | 1.7663 | 1.6684 | 1.7965 | 1.7325 |
| | | (−5.54) | (+1.71) | (−1.92) |
| 1 | 3.8391 | 3.7455 | 3.8712 | 3.8084 |
| | | (−2.44) | (+0.83) | (−0.80) |
| 2 | 6.4793 | 6.3874 | 6.5126 | 6.4500 |
| | | (−1.42) | (+0.51) | (−0.45) |
| 3 | 9.6317 | 9.5414 | 9.6664 | 9.6039 |
| | | (−0.94) | (+0.36) | (−0.29) |
| 4 | 13.2606 | 13.1727 | 13.2976 | 13.2352 |
| | | (−0.66) | (+0.28) | (−0.19) |
| 5 | 17.3411 | 17.2562 | 17.3810 | 17.3186 |
| | | (−0.49) | (+0.23) | (−0.13) |

### D. A Nonlinear Field Model

Let us also suggest a possible application of the broken SUSY in field theories. In analogy to the class of potentials characterized by the superpotential (62) we may consider a one-dimensional field theory defined by a Lagrangian density of the form

$$\mathcal{L} = \alpha \left| \frac{d\Phi}{dx} \right|^2 + \beta \Phi^{4d} + \gamma \Psi^\dagger \frac{d\Psi}{dx} + \eta \Phi^{2d-1} [\Psi^\dagger, \Psi] \quad (68)$$

which characterizes a one-dimensional Ginzburg-Landau field $\Phi(x)$ coupled to a fermionic field $\Psi(x)$. It is known [36] that the inverse correlation lengths for such models are given by the energy eigenvalues of the corresponding Hamiltonian (8). A quasi-classical estimate of the eigenvalues by (2) and (4) will thus give simple lower and upper bounds of correlation lengths, respectively. This might lead to some insights, in particular, to the nature of phase transitions of such field theories even in more than one dimension. A recent study of Kapuscik, Uzes and Barut [37] shows that an exact result of the nonlinear complex oscillator yields a spectrum very similar to the present quasi-classical result.

## V. CONCLUDING REMARKS

In this paper, we have derived from Feynman's path integral the SUSY counterparts of the Van Vleck-Pauli-Morette formula for the propagator and the Gutzwiller formula for the energy dependent Green function by the quasi-classical approximation. From the poles of the Green function in the quasi-classical limit, we have obtained the CBC formula (1) for exact SUSY and the new quantization condition (4) for broken SUSY. We have then pointed out a limited role of the broken SUSY formula in reproducing exact energy spectra and demonstrated by few examples that the broken SUSY formula, whenever applicable, is generally as good as and sometime better than the WKB formula. We have also observed that the interesting feature that in approximation the broken SUSY formula overestimates while the WKB formula underestimates. Because of this special feature, the approximation can be improved by taking an average of the results from the two formulas.

Unfortunately, the present approach stops short in answering the question as to why the CBC formula, and the broken SUSY formula in a limited extent, can yield exact spectra for shape-invariant potentials without the Langer-like modification. It has been argued that the exactness of the lowest stationary phase approximation may be assured if all the higher order corrections are zero. However, it is difficult to show in a general way that all the higher order contributions vanish. Furthermore, there are instances that the first few order corrections even to the WKB approximation vanish while the Langer-like replacement is still required. It has never been proven that the Langer-like correction to the WKB result is indeed stemming from the higher order terms in $\hbar$. The question remains to be answered.

As has been discussed in the end of Section III, the conditions at the turning points helps to discriminate good SUSY and broken SUSY. The antisymmetric condition $\phi(x_L) = -\phi(x_R)$ corresponds to a case where SUSY is good, while the symmetric condition $\phi(x_L) = \phi(x_R)$ implies broken SUSY. If $\phi^2(x)$ is symmetric about the axis of $x = (x_L + x_R)/2$, then the parity of the superpotential immediately dictates whether SUSY is broken or not [16]. It may be worth noticing that the fermionic phase $\varphi$ for full cycles precisely cancels the Morse index for each path, that is, $\varphi_k^{(i)} - \varphi_0^{(i)} = (\pi/2)\left(\nu_k^{(i)} - \nu_0^{(i)}\right)$, when SUSY is good. This is not true when SUSY is broken. Recently, it has also been pointed out that there is a close connection between the Maslov corrections and the parity of the underlying problem [35].

Finally, we have to touch upon shortcomigs of the quasi-classical approach. For a shape-invariant potential, the scalar function $\phi(x)$ can be found in a rather simple form. However, it is not easy to find the function $\phi(x)$ for an arbitrary potential $V(x)$. Even if we can get the function in an analytical form, the integration of the CBC formula (1) or the new formula (4) will be difficult. For instance, if $V(x) = \alpha x^n$, we have the Riccati equation [38] to solve

$$\phi'(x) + a\phi^2(x) = bx^n, \quad (69)$$

with $a = -\sqrt{2m}/\hbar$ and $b = -\alpha\sqrt{2m}/\hbar$, whose solution is given by $\phi(x) = a^{-1}u'(x)/u(x)$ with

$$u(x) = \sqrt{x}\, Z_{\frac{1}{n+2}} \left( \frac{2i\sqrt{ab}}{n+2}\, x^{(n+2)/2} \right) \quad (70)$$

where $Z_\nu(z)$ is a solution of Bessel's differential equation; for example, the Bessel function $J_\nu(z)$. Therefore, quantization cannot be done analytically by either the CBC formula or the broken SUSY formula (4). The WKB formula (2), however, leads to analytical results in this particular example. For a more general potential, we must solve the generalized Riccati equation,

$$\phi'(x) + a\phi^2(x) = V(x), \quad (71)$$

which is not an easy task. In this regard, neither the CBC nor broken SUSY formula surpasses the breadth of the WKB formula as an approximation formula even though there are occasions where the SUSY formulas give better estimations. Yet, we have to recognize that for many shape-invariant potentials the SUSY formulas have a mysterious power to yield exact results which the WKB rule fails in providing.

## ACKNOWLEDGMENTS

One of the authors (G.J.) was partially supported by the Deutsche Forschungsgemeinschaft which is gratefully acknowledged.

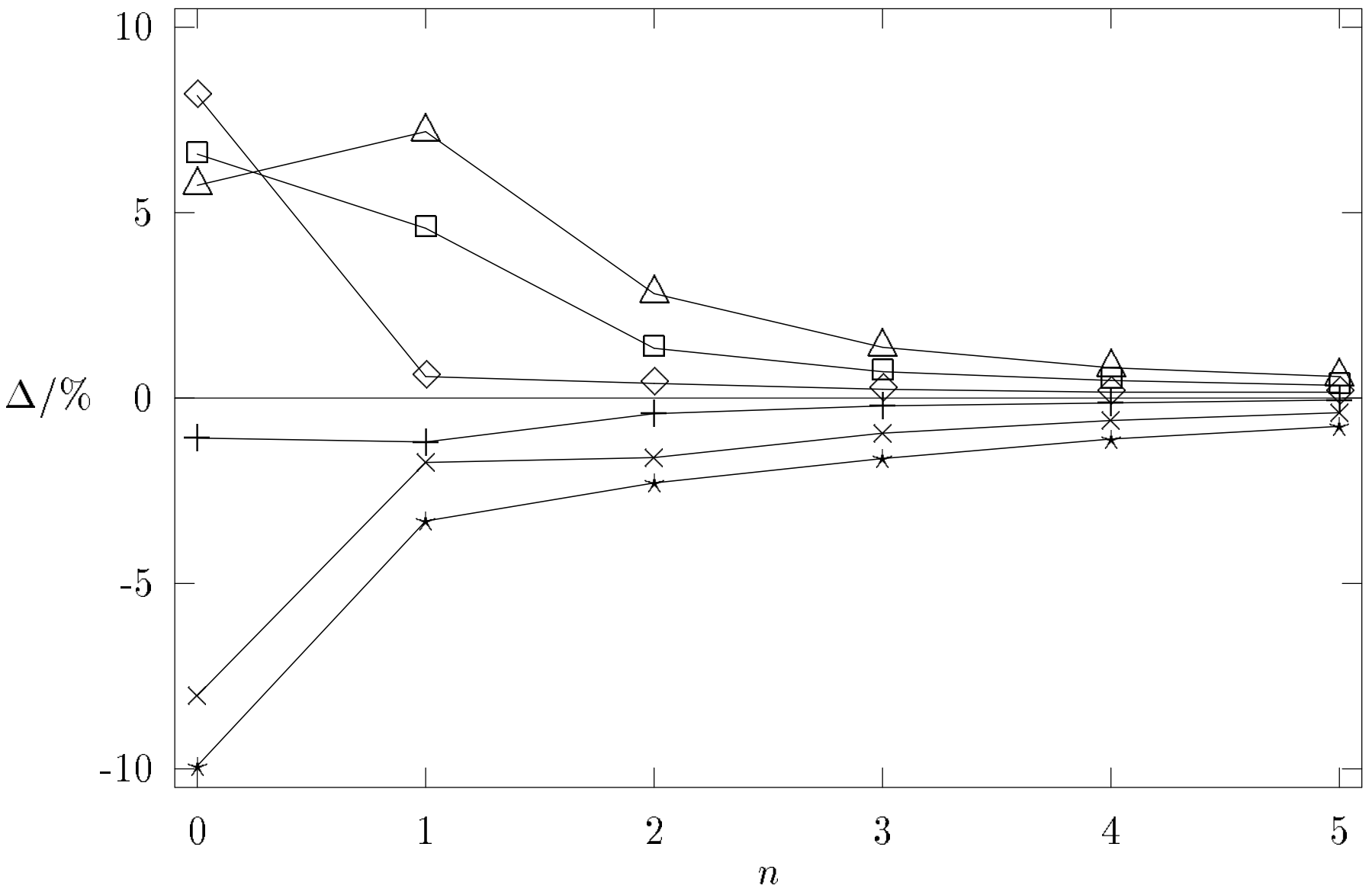

FIG. 1. Relative errors of WKB ($+, \times, \star$) and broken SUSY formula ($\diamond, \square, \triangle$) for the power potential (62) shifted by $\epsilon = 2$ and parameters $d = 1$ ($+, \diamond$), $d = 2$ ($\times, \square$) and $d = 3$ ($\star, \triangle$).

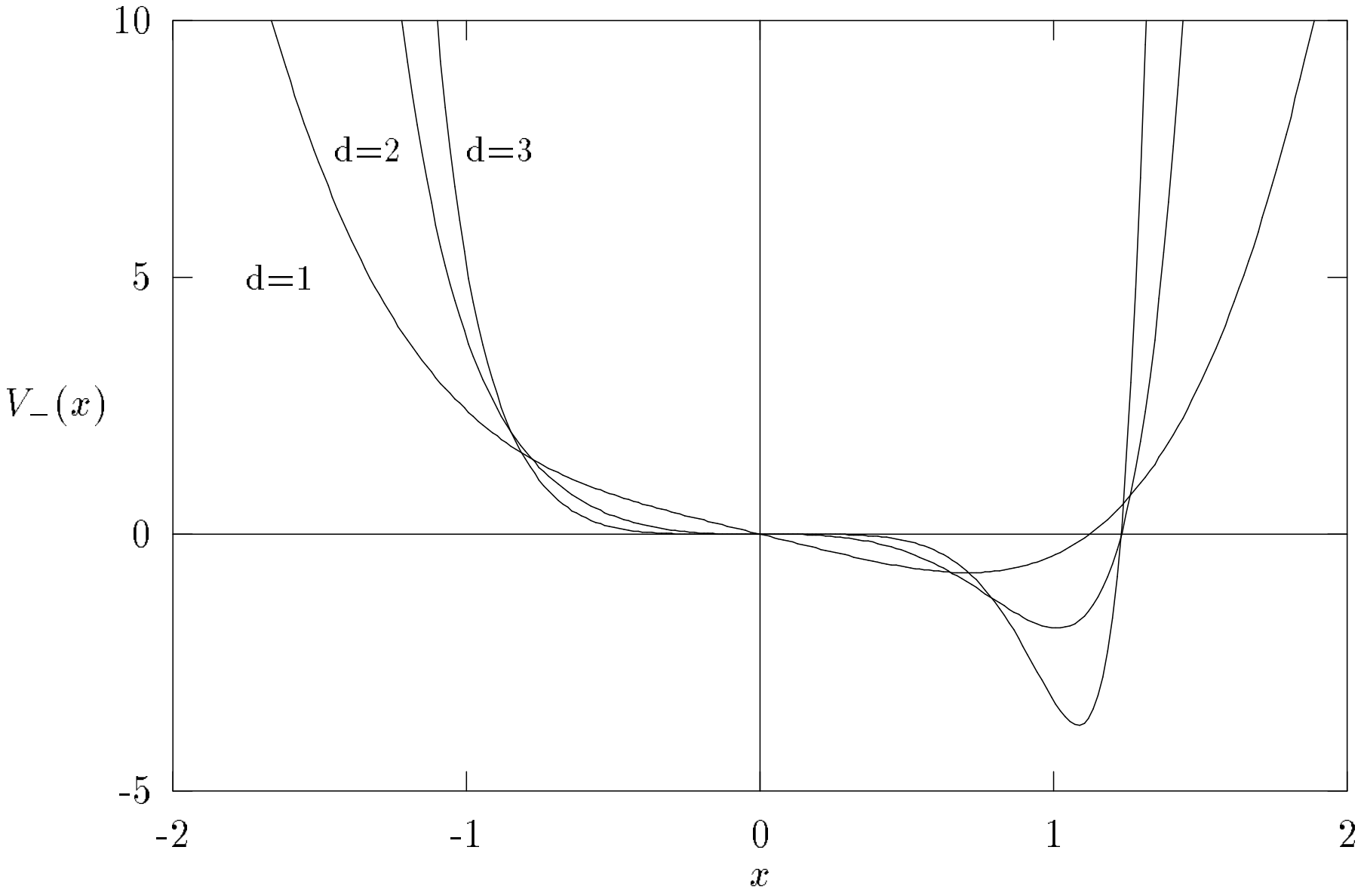

FIG. 2. The potential $V_{-}(x)$ in (62) for parameters $d = 1$, 2 and 3.

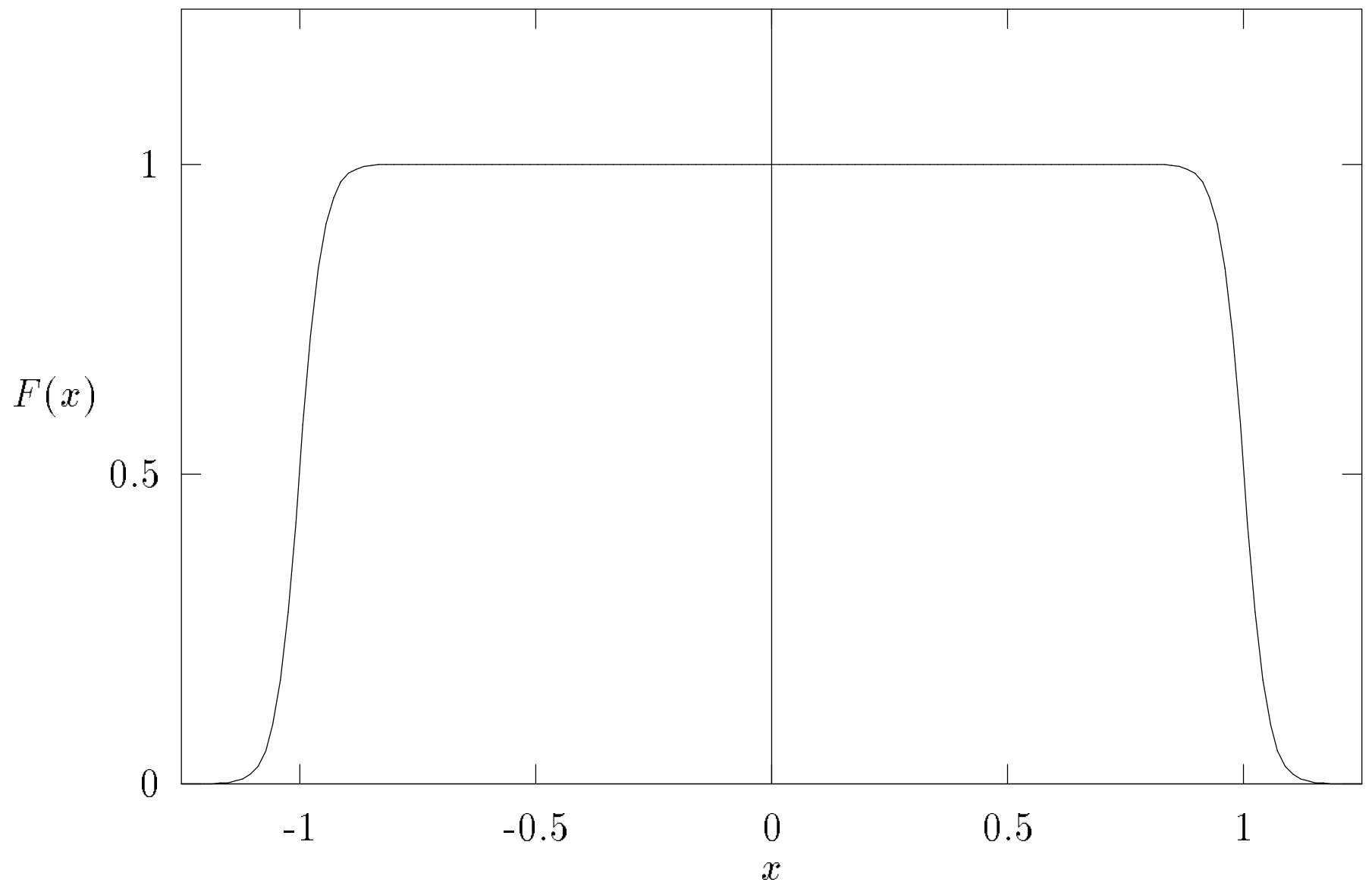

FIG. 3. The envelope function $F(x)$ defined in (65) for large $d$.

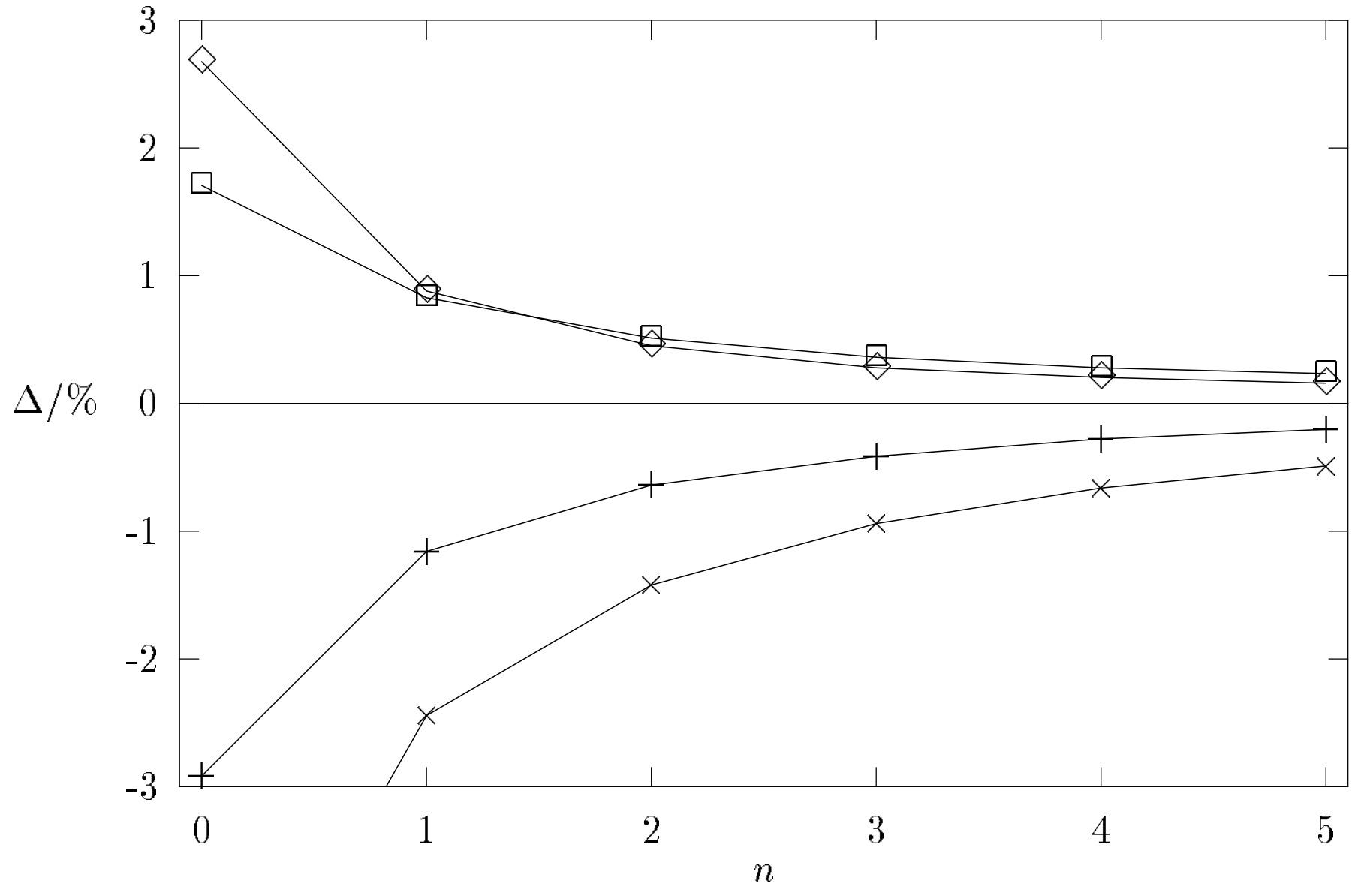

FIG. 4. Relative errors of WKB $(+,\times)$ and broken SUSY formula $(\diamond,\square)$ for superpotentials, $\phi(x) = \cosh\, x$ $(+,\diamond)$ and $\phi(x) = \exp(x^2/2)$ $(\times,\square)$.

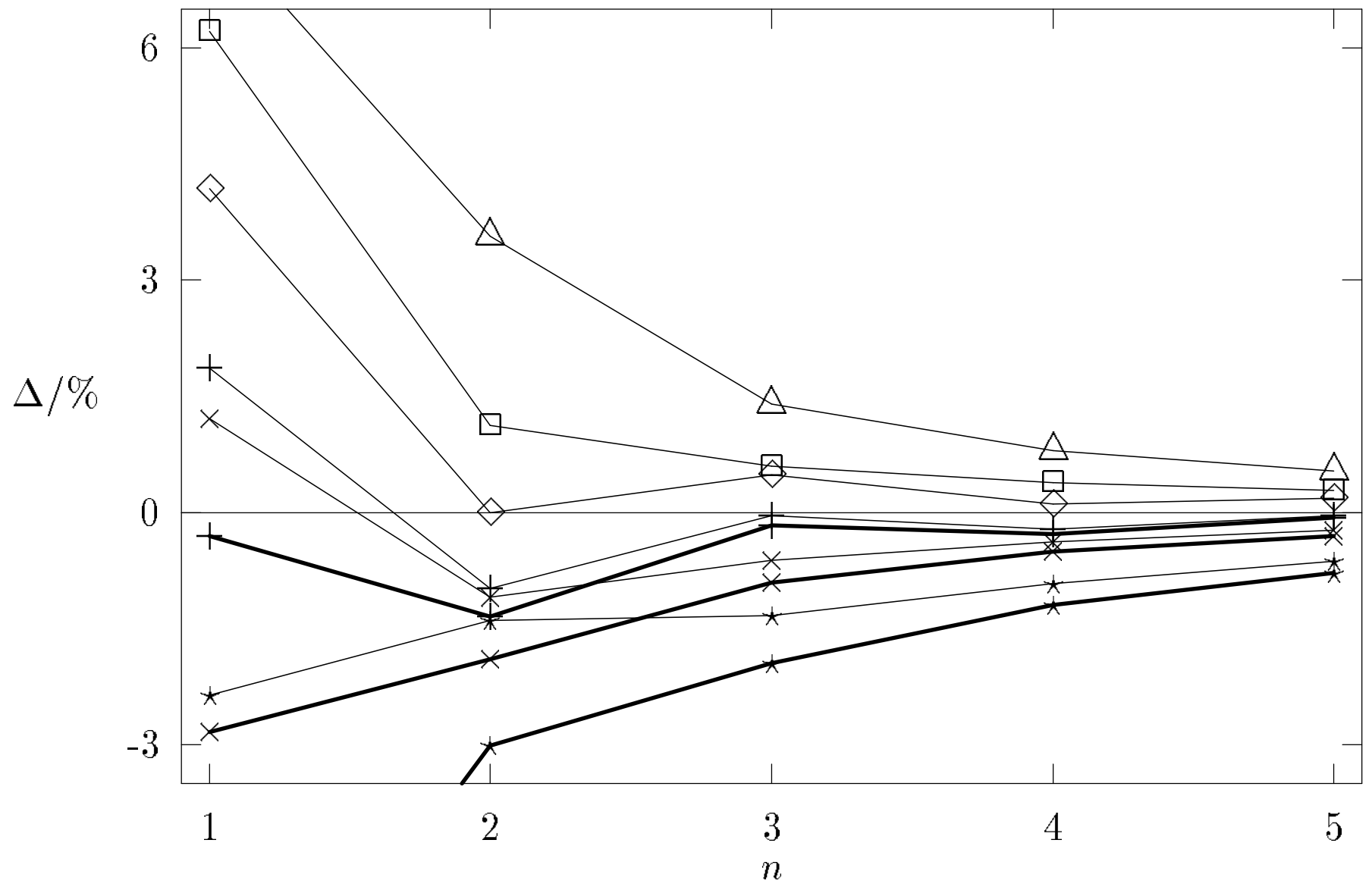

FIG. 5. Relative errors of WKB ($+, \times, \star$) and CBC ($\diamond, \square, \triangle$) for superpotentials, $\phi(x) = x|x|$ ($+, \diamond$), $\phi(x) = x^3$ ($\times, \square$) and $\phi(x) = x^5$ ($\star, \triangle$). The energy shift is $\epsilon = 2$.